%% file: main_arXiv.tex
\documentclass[a4paper]{article}

\usepackage{geometry}
\usepackage[utf8]{inputenc}
\usepackage[english]{babel}
\usepackage{authblk}
\usepackage{graphicx}
\usepackage{caption, subcaption}
\usepackage{multirow}
\usepackage{amsmath, amsfonts, amssymb}
\usepackage{braket}
\usepackage{hyperref}

\title{On Applying the Lackadaisical Quantum Walk Algorithm to Search for Multiple Solutions on Grids}

\author[1]{Jonathan H. A. de Carvalho\thanks{Corresponding author}}
\author[2]{Luciano S. de Souza}
\author[1]{Fernando M. de Paula Neto}
\author[2]{Tiago A. E. Ferreira}

\affil[1]{Centro de Inform\'atica, Universidade Federal de Pernambuco, Recife, Pernambuco, Brazil}
\affil[ ]{\textit{\{jhac,fernando\}@cin.ufpe.br}}
\affil[2]{Departamento de Estat\'istica e Inform\'atica, Universidade Federal Rural de Pernambuco, Recife, Pernambuco, Brazil}
\affil[ ]{\textit{\{luciano.serafim,tiago.espinola\}@ufrpe.br}}

\date{}

\begin{document}

\maketitle

\input{sections/00-abstract}

\input{sections/01-introduction}
\input{sections/02-lackadaisical_qw}
\input{sections/03-comparison_stops}
\input{sections/04-solutions_setups}
\input{sections/05-lqw_d-dim}
\input{sections/06-final_remarks}

\input{sections/07-acknowledgments}

%\bibliographystyle{ieeetr}
%\bibliography{references}

\input{main_arXiv.bbl}
\end{document}

%% file: sections/00-abstract.tex
\begin{abstract}

Quantum computing promises to improve the information processing power to levels unreachable by classical computation. Quantum walks are heading the development of quantum algorithms for searching information on graphs more efficiently than their classical counterparts. A quantum-walk-based algorithm standing out in the literature is the lackadaisical quantum walk. The lackadaisical quantum walk is an algorithm developed to search graph structures whose vertices have a self-loop of weight $l$. This paper addresses several issues related to applying the lackadaisical quantum walk to search for multiple solutions on grids successfully. Firstly, we show that only one of the two stopping conditions found in the literature is suitable for simulations. In the most discrepant case shown here, a stopping condition is prematurely satisfied at the step $T=288$ with a success probability $Pr=0.593276$, while the suitable condition captures the actual amplification that occurred until $T=409$ with $Pr=0.878178$. We also demonstrate that the final success probability depends on both the space density of solutions and the relative distance between solutions. For instance, we show here that decreases in the density of solutions can even take a success probability of $0.849178$ to $0.961896$. In contrast, increases in the relative distances can even take a success probability of $0.871665$ to $0.940301$. Furthermore, this work generalizes the lackadaisical quantum walk to search for multiple solutions on grids of arbitrary dimensions. In addition, we propose an optimal adjustment of the self-loop weight $l$ for such $d$-dimensional grids. It turns out other fits of $l$ found in the literature are particular cases. Our experiments demonstrate that successful searches for multiple solutions with higher than two dimensions are possible by achieving success probabilities such as $0.999979$, with the value of $l$ proposed here, where it would be $0.637346$, with the value of $l$ proposed in previous works. Finally, we observe a two-to-one relation between the steps of the lackadaisical quantum walk and Grover's algorithm, which requires modifications in the stopping condition. That modified stopping condition can escape intermediary fluctuations that would produce premature stops at $T=6$ with $Pr=0.000878$ where the system can evolve until $T=354$ with $Pr=0.99999$, as an example that we show here. In conclusion, this work deals with practical issues one should consider when applying the lackadaisical quantum walk, besides expanding the technique to a broader range of search problems.

\end{abstract}

%\begin{keyword}
%Quantum computing \sep Quantum walk \sep Lackadaisical quantum walk \sep Search algorithm \sep Spatial search.
%\end{keyword}

%% file: sections/01-introduction.tex
\section{Introduction}

Quantum computing is expected to demonstrate supremacy~\cite{preskill_quantum_supremacy} over classical computing through the exploration of inherently quantum phenomena such as superposition and entanglement~\cite{nielsen_QC}. The opportunities that emerge from the quantum realm have attracted significant efforts in research areas like information security~\cite{qu_qblockchain-framework_medical-IoT}, decision making~\cite{liu_qguided-expert-transition_decision-making}, artificial neural networks~\cite{situ_qgan_discrete-distribution}, and optimization~\cite{ruan_QAOA_constraints}. Regarding optimization, the scientific community is actively developing quantum or even quantum-inspired meta-heuristics of search, such as genetic algorithms and particle swarm optimization~\cite{acampora_HQGA_IBM-dev, fang_decentralized_QI-PSO, li_self-organizing_QI-PSO}. Searching is one of the tasks where quantum computing has the most known examples of speedup over classical counterparts, mainly due to the algorithm proposed by Grover~\cite{grover_algorithm_search}. Grover's algorithm can successfully search for a single element within a disordered array of $N$ items in $O(\sqrt{N})$ steps, which is a quadratic speedup over the classical analogs.

However, if the task is a spatial search, Benioff~\cite{benioff_robot_spatial_search} showed that a quantum robot using Grover's algorithm is no more efficient than a classical robot because both require $O(N\log{\sqrt{N}})$ steps to search 2-dimensional grids of size $\sqrt{N}$ x $\sqrt{N}$. It makes room for employing other techniques to search for information on physical regions modeled as connected graphs~\cite{aaronson_spatial_search}, also known as spatial search problems. Then, in pursuit of the speedup that Grover's algorithm failed to provide, researchers addressed that type of problem using quantum walks. First, Childs and Goldstone~\cite{childs_spatial_1st_CTQW} addressed a 2D spatial search problem using a continuous-time quantum walk but failed to provide substantial speedup. On the other hand, Ambainis et al.~\cite{ambainis_search_walk} proposed an algorithm capable of finding the solution in $O(\sqrt{N} \log N)$ steps using a discrete-time quantum walk. Childs and Goldstone~\cite{childs_spatial_CTQW_Dirac} showed later that a continuous-time model of quantum walks can also achieve this same speedup.

Over time, quantum walks for other graph structures have been developed~\cite{meyer_connectivity_runtimes, shenvi_QW_hypercube, wong_lackadaisical_complete_graph, tanaka_CTQW_Johnson-graphs, tanaka_DTQW_Johnson-graphs, qu_QW_star-graphs, qu_CTQWs-equivalence_irregular-graphs}, but the attempts to improve the search on 2D grids also continued. In particular, the lackadaisical quantum walk (LQW) developed in~\cite{wong_lackadaisical_2d} has been drawing attention because it improved the 2D spatial search by making a simple modification to the algorithm proposed in~\cite{ambainis_search_walk}. The modification was to attach a self-loop of weight $l$ at each vertex of the 2D grid. Adding a new degree of freedom to enable staying at the same position had already been studied for the quantum walk on the line~\cite{inui_lazy_QW_1D}, where analysis of time scaling~\cite{falcao_1D-lazy-QW_time-scaling}, entanglement entropy and temperature~\cite{tude_1D-lazy-QW_entanglement_entropy&temperature}, and decoherence~\cite{tude_1D-lazy-QW_decoherence} have been made recently. The LQW, in turn, added a weighted edge that points to the same vertex on the 2D grid. When the weight $l$ of the self-loop is optimally adjusted, the LQW can find the solution to the search problem in $O(\sqrt{N \log N})$ steps, which is an $O(\sqrt{\log N})$ improvement over that loopless version presented in~\cite{ambainis_search_walk}.

The LQW improvement was achieved by fitting the self-loop weight to $l=4/N$, where $N$ is the total number of vertices. This optimal value is only one instance of a general observation about the LQW searching vertex-transitive graphs with $m=1$ solutions. For these cases, the optimal value of $l$ equals the degree of the graph without loops divided by $N$~\cite{rhodes_LQW_vertex-transitive_graphs}. An analytical proof of this conjecture is given in~\cite{hoyer_proof_LQW_interpolated-QW} using the fact that the quantum interpolated walk can approximate the LQW. However, that conjecture about the adjustment of $l$ does not hold when the grid's number of solutions $m$ is higher than 1. Thus, another adjustment of $l$ is required. Nahimovs~\cite{nahimovs_lackadaisical} proposed two adjustments of $l$ for arbitrary placements of the solutions, both in the form $l=\frac{4(m-O(m))}{N}$. After that, Giri and Korepin~\cite{giri_LQW_runtime_m-solutions} showed that one of these $m$ solutions can be obtained with sufficiently high probability in $O(\sqrt{\frac{N}{m} \log \frac{N}{m}})$ steps. Saha et al.~\cite{saha_lackadaisical_block_solutions} showed that $l \approx \frac{4}{N(m + 1)}$ is the optimal value for the exceptional configuration of $m$ solutions arranged as a block of $\sqrt{m}$ x $\sqrt{m}$ within the grid.

\begin{figure}
    \centering
    \includegraphics[width=0.9\textwidth]{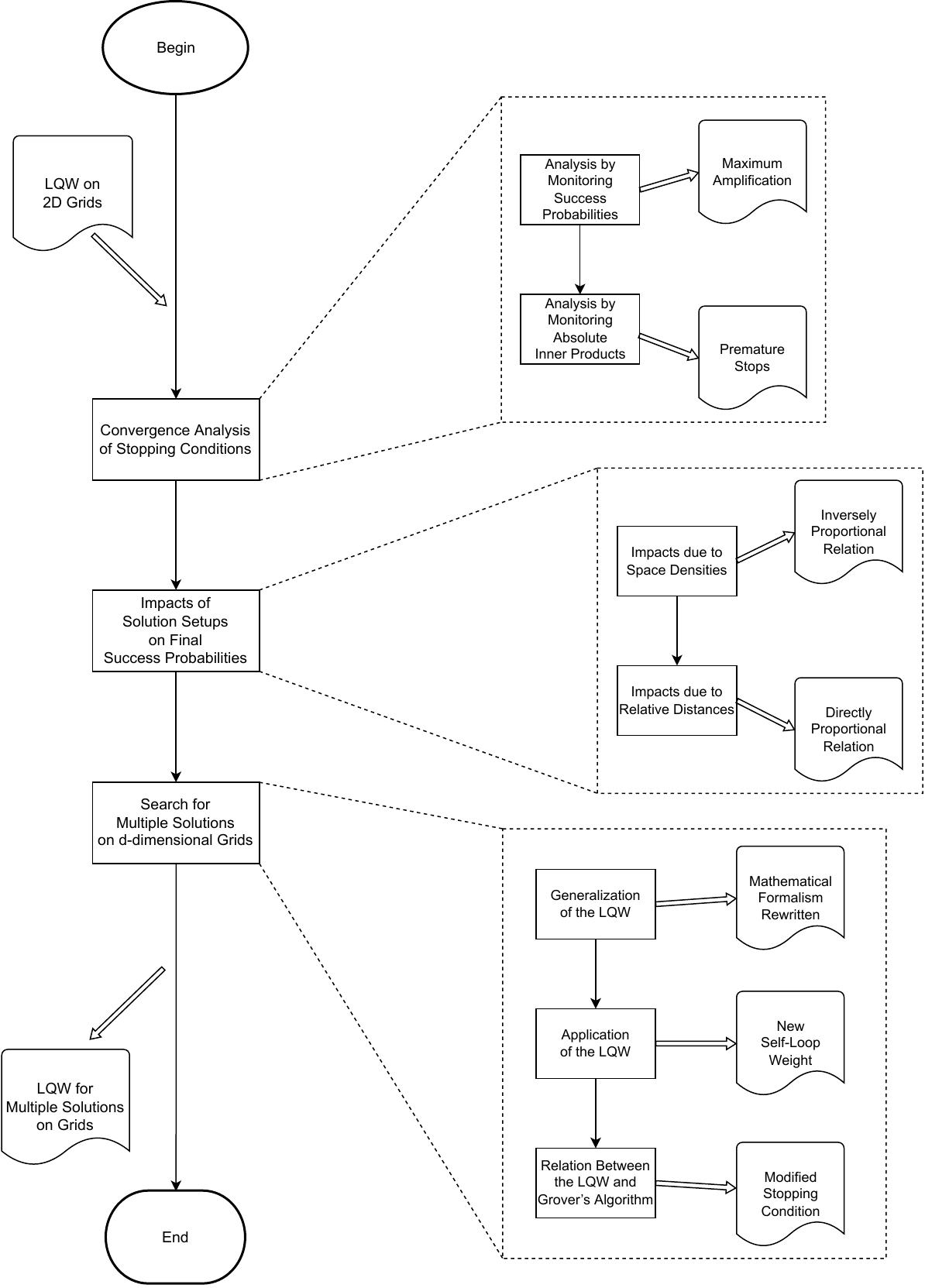}
    \caption{General flowchart representing the studies contemplated in this work, linked by connectors. Information that enters the process as previous knowledge or leaves the process as contributions are linked to the main flow through arrows. Each study is subdivided into minor activities in a more detailed view, where intermediary contributions are also presented.}
    \label{fig:general_flowchart}
\end{figure}

This paper is a solid continuation of the incipient conference paper presented in~\cite{jonathan_impacts_solutions_LQW}. Figure~\ref{fig:general_flowchart} shows a general flowchart of our complete research, which is an extensive experimental analysis of the LQW search algorithm. The main flow represents the studies performed here, linked by connectors. Previous knowledge and the knowledge discovered in this research are linked to the main flow by arrows. Each study is decomposed into minor activities with their intermediary contributions, which compose all the knowledge provided by this research to the scientific community.

From the previous efforts related to the search on 2D grids by the LQW algorithm, one of the ideas that can be learned is that the simulations should stop according to two interchangeable stopping conditions. However, we demonstrate through a convergence analysis that LQW simulations should stop according to the stopping condition used in~\cite{wong_lackadaisical_2d} only because that condition captures the maximum amplification of the system. In contrast, the stopping condition used in~\cite{nahimovs_lackadaisical} is satisfied prematurely, i.e., when the system can evolve even further. After, we investigate the impacts of solution setups on the final success probabilities achieved by the LQW algorithm. As a result, we show that the final success probability is inversely proportional to the space density of solutions and directly proportional to the relative distance between solutions. Those results have already been corroborated and extended in the literature. Nahimovs and Santos~\cite{nahimovs_LQW_2D-types} showed that the success probability is inversely proportional to the density of solutions not only on rectangular 2D grids but also on triangular and honeycomb 2D grids.

We finally consolidate the work by addressing the search for multiple solutions on $d$-dimensional grids. As the first step in this direction, we generalize the LQW algorithm to such a new scenario by rewriting its mathematical formalism. In this way, we enable the LQW to search for solutions with arbitrary dimensions. Retrieving higher than two quantities per solution is supposed to enlarge the spectrum of applications. For example, one application we glimpse is to use the LQW to search all weights and biases of artificial neural networks, inspired by works that applied a lackadaisical quantum walk for complete graphs~\cite{luciano_QW_ELM_RNA, luciano_QW_all-weights_RNA}. As quantum information routing by quantum walks can benefit from high dimensional cases~\cite{zhan_QW_state-transfer}, spatial information search by quantum walks can also. For example, spatial search by quantum walks on sufficiently high dimensions can allow the full $O(\sqrt{N})$ speedup, which is unfeasible in low dimensions~\cite{aaronson_spatial_search, childs_spatial_1st_CTQW}. Thus, the solution to the search problem may be retrieved even faster.

Proceeding to the application of the generalized LQW, a new optimal self-loop weight becomes necessary to achieve success in $d$-dimensional grids. We then propose a generalized adjustment for the value of $l$. It turns out that other fits reported in the literature~\cite{wong_lackadaisical_2d, rhodes_LQW_vertex-transitive_graphs, nahimovs_lackadaisical} are particular cases. The adjustment of $l$ proposed here has already been used in the literature as evidence to consider the number of solutions in the value of $l$ when searching hypercubes~\cite{luciano_LQW_hypercube_multiple-solutions}. Finally, we observe a two-to-one relation between the steps of the LQW and the steps of Grover's algorithm, which requires a modification in the stopping condition. In conclusion, this work addresses several practical issues that are critical to the successful application of the LQW algorithm when searching for multiple solutions on grids.

This paper is organized as follows. Section~\ref{section:lackadaisical_quantum_walk} presents the theoretical background about the task of search on 2D grids by the LQW algorithm. Here, the reader is expected to be familiar with the basics of quantum computing. If it is not the case, knowledge from the basic to the advanced levels can be obtained in~\cite{nielsen_QC, yanofsky_QC}. Section~\ref{section:comparison_stop_conditions} compares the different stopping conditions used in previous works. After, Section~\ref{section:solutions_setups} relates the impacts on the final success probability to both the space density of solutions and the relative distance between solutions. Section~\ref{section:lqw_d-dim_grids} includes generalizing the LQW to grids of arbitrary dimensions with multiple solutions, finding a new optimal value of $l$, and modifying the stopping condition to tolerate a meaningful kind of fluctuation. Finally, Section~\ref{section:final_remarks} presents concluding remarks.

%% file: sections/02-lackadaisical_qw.tex
\section{Search with the Lackadaisical Quantum Walk}
\label{section:lackadaisical_quantum_walk}

The classical random walk is a probabilistic movement in which a particle jumps to its adjacent positions based on the outcome of a non-biased random variable at each step~\cite{portugal_walks}. Generally, the random variable is a fair coin with one degree of freedom for each possible direction of movement in the space at hand. This simple concept can also support sophisticated approaches to practical problems, such as random-walk-based recommendations of potential lenders in peer-to-peer lending~\cite{zhang_random-walk_P2P-lending}.

The quantum walk, in turn, is a generalized concept compared to the classical random walk. That high-level idea of conditioned movements remains, but quantum operations are responsible for evolving the system. In this context, quantum properties such as interference and superposition allow the quantum walk to spread quadratically faster than the classical one~\cite{portugal_walks}. This advantage, therefore, can be used to search spatial regions more efficiently~\cite{wong_intro_search_walks}.

\subsection{Spatial Search with a Quantum Walk}

Ambainis et al.~\cite{ambainis_search_walk} proposed a quantum walk algorithm to search a single vertex, also called the marked vertex, in the 2-dimensional grid of $L \times L = N$ vertices. In that work, the process evolved on the Hilbert space $\mathcal{H} = \mathcal{H}_{C} \otimes \mathcal{H}_{P}$, where $\mathcal{H}_{C}$ is the 4-dimensional coin space, spanned by $\{ \ket{\uparrow}, \ket{\downarrow}, \ket{\leftarrow}, \ket{\rightarrow} \}$, and $\mathcal{H}_{P}$ represents the $N$-dimensional space of positions, spanned by $\{\ket{x,y} : x,y \in [0,\ldots, L - 1]\}$.

Firstly, the coin toss is accomplished by the operator $C$ presented in Equation~\ref{eq:coin_operator_ambainis}, which combines the coin operators $C_0$ and $C_1$ in such a way that $C_1$ is applied only to the marked state $\ket{v}$, while $C_0$ is applied to the others. This idea of different evolution regimes for marked and unmarked vertices was introduced in~\cite{shenvi_QW_hypercube}. Particularly, Ambainis et al.~\cite{ambainis_search_walk} defined $C_0$ as the Grover diffusion coin: $C_0 = 2\ket{s}\bra{s} - I_4$, where $\ket{s} = \frac{1}{2} (\ket{\uparrow} + \ket{\downarrow} + \ket{\leftarrow} + \ket{\rightarrow})$ and $I_4$ denotes the $4$-dimensional identity operator. Finally, $C_1$ was defined as $-I_4$. Thus, $-I_4$ is applied to the marked state, while the Grover diffusion coin is applied to the others.

\begin{equation}
\label{eq:coin_operator_ambainis}
C = C_0 \otimes (I_4 - \ket{v}\bra{v}) + C_1 \otimes \ket{v}\bra{v}
\end{equation}

Then, the flip-flop shift operator $S_{ff}$ is applied to move the quantum particle while inverting the coin state, as presented in Equation~\ref{eq:flip_flop_shift_operator}. This shift works $\mod \sqrt{N} = L$ because the grid has periodic boundary conditions. Finally, the quantum walk is a repeated application of the operator $U = S_{ff} \cdot C$ to the quantum system $\ket{\psi}$, which begins in the state $\ket{\psi(0)} = \frac{1}{\sqrt{N}} \sum_{x,y=0}^{\sqrt{N}-1} \ket{s} \otimes \ket{x,y}$.

\begin{equation}
\label{eq:flip_flop_shift_operator}
\begin{aligned}
S_{ff} \ket{\rightarrow}\ket{x,y} = \ket{\leftarrow}\ket{x+1,y}
\\
S_{ff} \ket{\leftarrow}\ket{x,y} = \ket{\rightarrow}\ket{x-1,y}
\\
S_{ff} \ket{\uparrow}\ket{x,y} = \ket{\downarrow}\ket{x,y+1}
\\
S_{ff} \ket{\downarrow}\ket{x,y} = \ket{\uparrow}\ket{x,y-1}
\\\\
\end{aligned}
\end{equation}

As a result, the marked vertex can be obtained at the measurement with a probability $O(1/\log N)$ after $T = O(\sqrt{N \log N})$ steps. To achieve a success probability near to 1, amplitude amplification~\cite{brassard_amplitude_amplification} was applied, which implied additional $O(\sqrt{\log N})$ steps. Hence, the total running time of this quantum-walk-based search algorithm is $O(\sqrt{N} \log N)$.

\subsection{Improved Running Time by the Lackadaisical Quantum Walk}

The LQW search algorithm~\cite{wong_lackadaisical_2d} is an approach strictly based on the algorithm designed in~\cite{ambainis_search_walk} that we just discussed. The main modification is to attach a self-loop of weight $l$ at each vertex of the 2D grid, which implies other changes in the loopless technique. First, $\mathcal{H}_C$ is spanned now by $\{ \ket{\uparrow}, \ket{\downarrow}, \ket{\leftarrow}, \ket{\rightarrow}, \ket{\circlearrowleft} \}$ because of the new degree of freedom. However, no changes are required for $\mathcal{H}_P$.

Regarding the coin operator, $C_0$ was defined as the Grover diffusion coin for weighted graphs~\cite{wong_coin_weighted_graphs}, so $C_0 = 2 \ket{s_c}\bra{s_c} - I_5$, where $\ket{s_c}$ is the non-uniform distribution presented in Equation~\ref{eq:weighted_coin_states_2d}, and $I_5$ denotes the 5-dimensional identity operator. Also, better results were found when $C_1 = -C_0$, outperforming that choice of $C_1 = -I$ used in~\cite{ambainis_search_walk}. About the shift operator $S_{ff}$, it works like an identity operator when applied to $\ket{\circlearrowleft}\ket{x,y}$. Finally, the quantum system $\ket{\psi}$ begins in a uniform distribution between all vertices with their edges in the weighted superposition $\ket{s_c}$ presented in Equation~\ref{eq:weighted_coin_states_2d} instead of the uniform $\ket{s}$.

\begin{equation}
\label{eq:weighted_coin_states_2d}
    \ket{s_c} = \frac{1}{\sqrt{4+l}} (\ket{\uparrow} + \ket{\downarrow} + \ket{\leftarrow} + \ket{\rightarrow} + \sqrt{l}\ket{\circlearrowleft})
\end{equation}

As a result, the LQW with $l=4/N$ finds the marked vertex with a success probability close to 1 after $T=O(\sqrt{N \log N})$ steps. It is an $O(\sqrt{\log N})$ improvement over the loopless algorithm. More sophisticated approaches can also achieve this improvement in the running time, like in~\cite{portugal_faster_search_hamiltonians}, but the LQW is a significantly simpler and equally capable technique. Moreover, the success probability converges closer and closer to 1 if the number of vertices $N$ increases when using that optimal $l$.

Those numerical results reported in~\cite{wong_lackadaisical_2d} were found by simulations that stopped when the first peak in the success probability occurred. For that, the stopping condition monitored the success probability at each step. When the current value was smaller than the immediately previous one for the first time, the simulation stopped, and this immediately previous result was reported as the maximum found.

\subsection{Lackadaisical Quantum Walk with Multiple Solutions}

If there are multiple marked vertices, i.e., multiple solutions in the search space, the LQW results for the case with only one solution do not hold. Such cases with multiple solutions require new optimal choices of the self-loop weight $l$. In this way, Nahimovs~\cite{nahimovs_lackadaisical} optimally adjusted the value of $l$ for the cases where $m$ solutions are randomly sampled within the 2D grid.

This multiple-solution adjustment of $l$ occurred by searching for new optimal values in the form $l = \frac{4}{N} \cdot a$, where $a$ is a modifiable multiplicative factor. Thus, $l$ was adjusted as a factor of the optimal value for $m=1$ reported in~\cite{wong_lackadaisical_2d}, which was $l=4/N$. As a result, two adjustments were proposed: $l = \frac{4m}{N}$, for small values of $m$, and $l=\frac{4(m-\sqrt{m})}{N}$, for large values of $m$. To find these optimal values of $l$, the $m$ solutions were arranged following the $M_m$ set presented in Equation~\ref{eq:nahimovs_set_solutions}. However, random placements of solutions yielded similar results.

\begin{equation}
\label{eq:nahimovs_set_solutions}
    M_m = \{ (0,10i) \ | \ i \in [0,m-1] \}
\end{equation}

Regarding the simulations performed in~\cite{nahimovs_lackadaisical}, a different stopping condition was used rather than monitoring the success probability at each step. Alternatively, the inner product $|\langle \psi (t)|\psi(0) \rangle |$ was monitored until its minimum was achieved, so the simulation stopped when this inner product became close to 0 in absolute value for the first time.

%% file: sections/03-comparison_stops.tex
\section{Comparison between Different Stopping Conditions}
\label{section:comparison_stop_conditions}

It is possible to find two stopping conditions in the literature regarding the LQW simulations. One of them is to monitor the success probability until its maximum is achieved~\cite{wong_lackadaisical_2d}, where ``success probability" refers to the probability of measuring a marked vertex (a solution to the search problem). Here, we name this stopping condition as ``Marked Vertices". The other stopping condition found in the literature is to monitor the inner product $|\langle \psi(t) | \psi(0) \rangle|$ until its minimum is achieved~\cite{nahimovs_lackadaisical}. However, as the convergence of those stopping conditions has not been compared, their interchangeability became an open question. Therefore, before conducting further experimental analysis of the LQW search algorithm, we verified whether those conditions converge to the same points from equal initial settings.

We conducted experiments using a setup equal to the one used in~\cite{nahimovs_lackadaisical}, i.e., a grid of $200$ x $200$ vertices with the $m$ solutions following the $M_m$ set. Constrained by this $M_m$ scheme, up to $20$ solutions can be placed in that space, since $M_{20} = \{(0,0), (0, 10), \ldots,(0, 180), (0, 190)\}$. Placing more than $20$ solutions in the $200$ x $200$ grid would require an organization other than the $M_m$ set so that the grid limits would not be extrapolated. However, as we used the $M_m$ scheme, we made experiments with $1$, $5$, $10$, $15$ and $20$ solutions in the grid.

As a result, the stopping conditions converged to the same points for $l = \frac{4m}{N}$, suggesting that the stopping conditions are equivalent. However, it is not the case for $l=\frac{4(m - \sqrt{m})}{N}$. Table~\ref{table:convergence-marked-vs-inner} contrasts the results obtained for $l=\frac{4(m - \sqrt{m})}{N}$ when monitoring both the marked vertices and the inner product $|\langle \psi (t)|\psi(0) \rangle |$. As can be seen, the results tend to converge to the same points as $m$ increases. Nevertheless, the conditions were not equivalent because each was satisfied at a different step $T$, which implied different final success probabilities $Pr$. Also, monitoring the inner product $|\langle \psi (t)|\psi(0) \rangle |$ generated lower success probabilities in all cases.

\begin{table}
    \centering
    \caption{Convergence step $T$ and final success probability $Pr$, as the number of solutions $m$ increases, for the different stopping conditions used in previous works with $l=\frac{4(m - \sqrt{m})}{N}$.}
    \label{table:convergence-marked-vs-inner}
    \begin{tabular}{c|cc|cc}
        \hline
        \multirow{3}{*}{$m$} & \multicolumn{4}{|c}{Stopping Conditions} \\ \cline{2-5}
        & \multicolumn{2}{c|}{Marked Vertices} & \multicolumn{2}{c}{$|\langle \psi (t)|\psi(0) \rangle |$} \\
        \cline{2-5}
        & $T$ & $Pr$ & $T$ & $Pr$ \\ \hline
        1 & 399 & 0.140828 & 420 & 0.138489 \\
        5 & 409 & 0.878178 & 288 & 0.593276 \\
        10 & 297 & 0.867440 & 249 & 0.704010 \\
        15 & 290 & 0.835395 & 254 & 0.747045 \\
        20 & 288 & 0.818635 & 268 & 0.778724 \\
        \hline
    \end{tabular}
\end{table}

\begin{figure}
    \centering
    \includegraphics[width=0.6\textwidth]{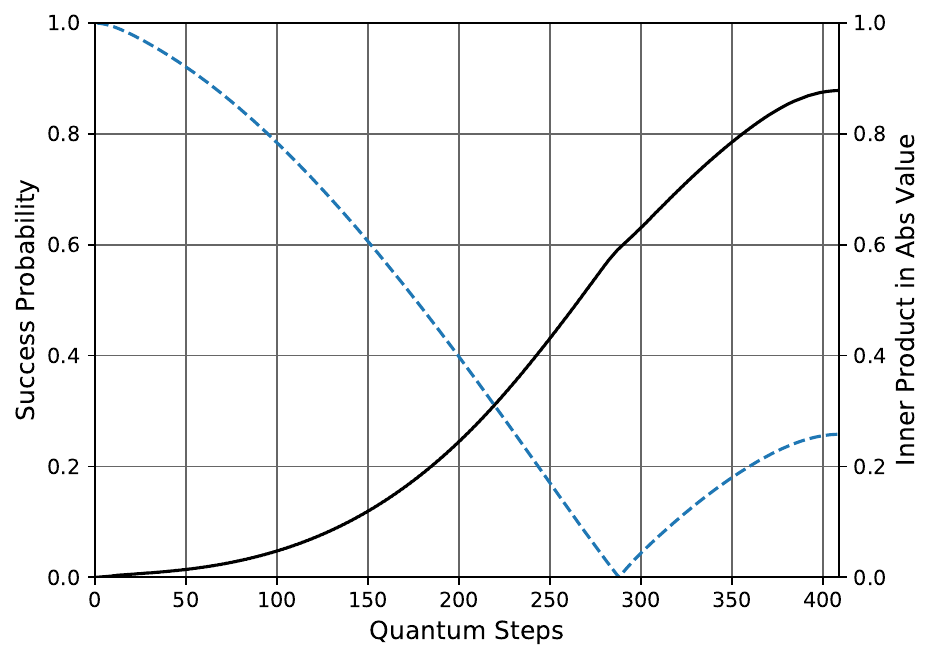}
    \caption{System evolution step by step until the condition that monitors the marked vertices is satisfied. The solid black line monitors that condition, while the dashed blue line monitors the inner product in absolute value.}
    \label{fig:marked_vs_abs-inner}
\end{figure}

Step by step, the system evolution was stored to investigate the divergence carefully. Figure~\ref{fig:marked_vs_abs-inner} shows the evolution of the $m=5$ case, which has the most significant discrepancy in Table~\ref{table:convergence-marked-vs-inner}. The solid black line represents the condition that monitors the marked vertices, while the dashed blue line represents the one that monitors the inner product $|\langle \psi (t)|\psi(0) \rangle |$.

As reported in Table~\ref{table:convergence-marked-vs-inner}, the condition that monitors the inner product in absolute value is satisfied prematurely at the step $T = 288$. It is said premature because the success probability continues increasing until $T = 409$. After the step $T = 288$, though, the curves have a similar growth damping, which raised a question about monitoring the real value of the inner product rather than its absolute value.

\begin{figure}
    \centering
    \includegraphics[width=0.6\textwidth]{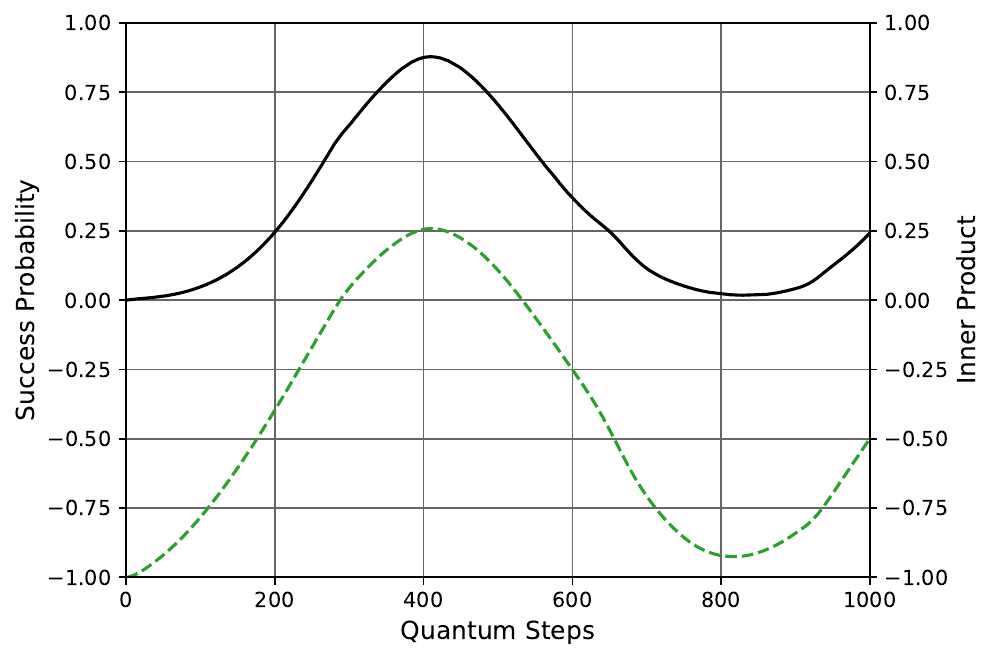}
    \caption{System evolution step by step during $1000$ steps. The solid black line monitors the marked vertices, while the dashed green line monitors the inner product without calculating its absolute value.}
    \label{fig:marked_vs_inner}
\end{figure}

Figure~\ref{fig:marked_vs_inner} shows the system evolution during $1000$ steps. As before, the monitoring of the marked vertices is represented by the solid black line. At this time, the inner product is monitored without calculating its absolute value, represented by the dashed green line. Note that both curves have the same behavior. Therefore, it is possible to conclude that the stopping conditions used in previous works are equivalent if, and only if, the inner product is considered without calculating its absolute value. Otherwise, only the condition that monitors the marked vertices leads to the real amplitude amplification achieved by the LQW search algorithm.

In this manner, all results discussed in this work from now on were found using the stopping condition that monitors the marked vertices. Since the objective is to measure the quantum system when the maximum amplification in the success probability occurs, monitoring the marked vertices is the most natural choice to define when the simulation should stop.

%% file: sections/04-solutions_setups.tex
\section{Solution Setups Affecting the Success Probability}
\label{section:solutions_setups}

After choosing the stopping condition properly in Section~\ref{section:comparison_stop_conditions}, we now move forward to addressing factors that affect the final success probability achieved by the technique. Previous works have already demonstrated the considerable dependence of the LQW algorithm on the self-loop weight $l$~\cite{wong_lackadaisical_2d, nahimovs_lackadaisical}. Expanding those analyses, we address the density of solutions and the relative distance between solutions.

\subsection{Previous Evaluations of Solution Densities and a Complementary Experiment}

The experiments performed by Wong~\cite{wong_lackadaisical_2d} and Nahimovs~\cite{nahimovs_lackadaisical} can reveal some dependence between the success probability and the space density of solutions $\rho$, where $\rho = \frac{m}{N}$. However, such works did not link the density of solutions and the success probability achieved at the end of the simulation. Here, we briefly discuss these previous experiments identifying the impacts of $\rho$. Finally, a complementary experiment was performed.

Firstly, Wong~\cite{wong_lackadaisical_2d} investigated the impacts of adding more unmarked vertices in a grid with only one solution. That experiment evaluated how decreases in the density of solutions could affect the final success probability. As a result, the success probability tends to improve, even though some disturbed behavior for the first values of $N$ exists.

After that, Nahimovs~\cite{nahimovs_lackadaisical} inserted more and more solutions in the $200$ x $200$ grid when adjusting the value of $l$ for multiple marked vertices. Since the grid size was fixed, that experiment increased the density of solutions with each new vertex being marked. However, the probability of measuring a marked vertex was smaller when $m$ increased. It would be a counter-intuitive idea in a classical world.

It is possible to comprehend the dependence between the total number of vertices $N$, the number of solutions $m$, and the final success probability observing Grover's algorithm~\cite{portugal_walks}. Consider $\ket{\omega}$ as the state where the total energy of the quantum system is equally distributed only between the marked states, so the success probability is $1$. The goal of Grover's algorithm is to rotate the system's state $\ket{\psi}$ to get as close to $\ket{\omega}$ as possible. However, this is an iterative process in which $\ket{\psi}$ rotates at each step by an angle $\theta$. As $\theta$ is inversely proportional to $N$, increasing $N$ implies more steps $T$, but these fine rotations turn $\ket{\psi}$ closer to $\ket{\omega}$ as $N$ increases, explaining the results of Wong~\cite{wong_lackadaisical_2d}. As $\theta$ is proportional to $m$, increasing $m$ implies fewer steps $T$ if $N \gg m$, although $\ket{\psi}$ gets less close to $\ket{\omega}$ at the end, explaining the smaller success probabilities in the experiments of Nahimovs~\cite{nahimovs_lackadaisical} while $m$ increased.

Thus, these previous experiments suggest that the success probability is inversely proportional to the density of solutions within the grid due to having a more refined or less refined angle $\theta$ in Grover's rotations, as explained. In this work, a complementary experiment was made to fill the gap not addressed by those previous works: decreasing the density of solutions, like in~\cite{wong_lackadaisical_2d}, in a grid with multiple marked vertices, like in~\cite{nahimovs_lackadaisical}. While we conducted this experiment, we also searched for the optimal value of $l$ in the form $l=\frac{4}{N}a$, like in~\cite{nahimovs_lackadaisical} again.

\begin{figure}
    \centering
    \includegraphics[width=0.6\textwidth]{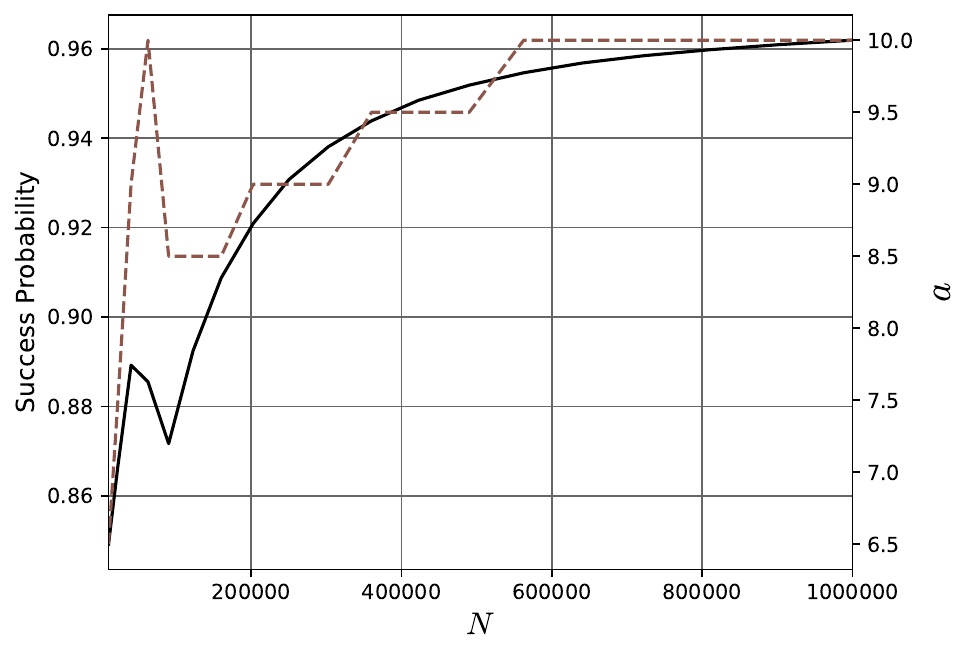}
    \caption{Peaks in the success probability, represented by the solid black line, and respective optimal values of $a$, represented by the dashed brown line, both as functions of $N$, with the $m=10$ solutions located in the 2D grid according to the $M_{10}$ set.}
    \label{fig:eval_density_10m}
\end{figure}

Figure~\ref{fig:eval_density_10m} shows the peaks in the success probability, represented by the solid black line, and the optimal $a$ values that generated these peaks, represented by the dashed brown line, both as functions of $N$. The density of solutions decreased in this case because $m$ was always equal to 10, while $N$ increased by adding unmarked vertices. The optimal values of $a$ were searched with a step size of $0.5$, and $N$ varied from $10^4$ to $10^6$ with the $m=10$ solutions following the $M_{10}$ set.

Again, there is a disturbed behavior for the first values of $N$, like in~\cite{wong_lackadaisical_2d}. Afterward, the success probability tends to 1, and the optimal value of $a$ tends to the number of solutions $m=10$. This experimental result suggests that the construction $l=\frac{4(m-O(m))}{N}$ proposed in~\cite{nahimovs_lackadaisical} is a way of adjusting for the cases where the density of solutions is not small enough. In the best cases of solutions density, $a$ equals $m$ and, consequently, $l=\frac{4m}{N}$.

\subsection{A New Set of Solutions Increasing Relative Distances}

In the last experiment, the $m=10$ solutions were located always at the points $\{(0,0), (0,10), \ldots,\allowbreak (0,90)\}$, following the $M_{10}$ set. That solutions distribution did not take advantage of the gradual increment in the total number of vertices $N$. If the solutions were located far from each other, it would be possible to continue evaluating how the success probability depends on the density of solutions but also on the relative distance between solutions.

Thus, we propose an alternative to the $M_m$ set that is the $P_{L, m}$ set presented in Equation~\ref{eq:new_set_solutions}. Following this new set, the $m$ solutions are located depending on the number of vertices in each dimension $L$ so that the grid size is better used. For example, $m=10$ solutions on the $200$ x $200$ grid would be located at the points $\{(0,0), (20,20), \ldots, (180,180)\}$, following the $P_{200,10}$ set. Hence, the solutions are farther apart using the $P_{L, m}$ set than the $M_m$ set.

\begin{equation}
\label{eq:new_set_solutions}
    P_{L,m} = \Bigg\{ \bigg( \Big\lfloor \frac{L}{m} \Big\rfloor i, \Big\lfloor \frac{L}{m} \Big\rfloor i \bigg) \ \ \Big| \ \ i \in [0, m-1] \Bigg\}
\end{equation}

Then, our complementary experiment that evaluated decreases in the density of solutions with $m=10$ solutions was redone, but using the $P_{L, m}$ set this time to localize the solutions farther from each other. The results obtained with this new set of solutions are contrasted in Table~\ref{table:comparison_sets_solutions} with the ones obtained previously, which used the $M_m$ set. The success probabilities in Table~\ref{table:comparison_sets_solutions} for the $M_m$ set are precisely the ones already presented in Figure~\ref{fig:eval_density_10m} but in terms of $L$ now because $L$ is the variable used to define the $P_{L, m}$ set. It is worth reminding that $L^2 = {N}$.

\begin{table}
    \centering
    \caption{Number of steps $T$ and final success probability $Pr$ as $L$ increases for different sets of $m=10$ solutions.}
    \label{table:comparison_sets_solutions}
    \begin{tabular}{c|cc|cc}
        \hline
        \multirow{2}{*}{$L$} & \multicolumn{2}{c|}{$M_{m=10}$} & \multicolumn{2}{c}{$P_{L, m=10}$} \\
        \cline{2-5}
        & $T$ & $Pr$ & $T$ & $Pr$ \\
        \hline 
        100 & 147 & 0.849178 & 109 & 0.902339 \\
        200 & 293 & 0.889219 & 223 & 0.927680 \\
        300	& 511 &	0.871665 & 342 & 0.940301 \\
        400 & 747 &	0.908749 & 460 & 0.948348 \\
        500 & 965 &	0.930714 & 581 & 0.953927 \\
        600 & 1181 & 0.943863 &	700 & 0.958288 \\
        700 & 1407 & 0.951843 &	822 & 0.961646 \\
        800 & 1623 & 0.956787 &	941 & 0.964317 \\
        900 & 1857 & 0.959761 &	1063 & 0.966613 \\
        1000 & 2097 & 0.961896 & 1187 &	0.968522 \\
        \hline
    \end{tabular}
\end{table}

For all values of $L$, the set of solutions $P_{L, m}$ generated better results because the success probability was higher and with fewer steps. Besides this, the disturbed behavior for the first values of $L$ did not appear in the results with the new set of solutions. Finally, it is possible to conclude from these numerical results that the success probability is directly proportional to the relative distance between solutions.

Regardless of whether or not a disturbed behavior exists for the first values of $L$, the success probability had an asymptotic and growing behavior for higher values of $L$ in all previous cases discussed here. However, that is not true for all values of $m$. Table~\ref{table:disturbed_behavior_m-view} shows the same investigation of decreases in the density of solutions with the $P_{L, m}$ set again, but for $m = \{3,4,5\}$, and not for $m=10$ as before.

\begin{table}
    \centering
    \caption{Values of $m$ that do not have asymptotic and growing behaviors for the success probability as the density of solutions decreases.}
    \label{table:disturbed_behavior_m-view}
    \begin{tabular}{c|ccc}
        \hline
        \multirow{2}{*}{$L$} & \multicolumn{3}{c}{$Pr$} \\
        \cline{2-4}
        & $m=3$ & $m=4$ & $m=5$ \\
        \hline
        100 & 0.991433 & 0.986119 & 0.981772 \\
        200 & 0.988165 & 0.992391 & 0.990397 \\
        300 & 0.985744 & 0.993451 & 0.992697 \\
        400 & 0.984221 & 0.993206 & 0.993754 \\
        500 & 0.983418 & 0.992604 & 0.994283 \\
        600 & 0.983100 & 0.991933 & 0.994585 \\
        700 & 0.983081 & 0.991282 & 0.994717 \\
        800 & 0.983252 & 0.990683 & 0.994677 \\
        900 & 0.983548 & 0.990138 & 0.994557 \\
        1000 & 0.983927 & 0.989644 & 0.994392 \\
        \hline
    \end{tabular}
\end{table}

The qualitative behaviors found for these $m$ values are not equal to the behaviors for both $m=1$, as reported in~\cite{wong_lackadaisical_2d}, and $m=10$, as shown in Figure~\ref{fig:eval_density_10m} and Table~\ref{table:comparison_sets_solutions}. In those $m=1$ and $m=10$ cases, a disturbed behavior existed during a transition from small to high values of $L$, and then the success probability improved continuously. However, $m = \{3,4,5\}$ can be seen as a kind of transition from small to high values from the perspective of the number of solutions $m$. It suggests that the asymptotic and growing behavior for the success probability as the density of solutions decreases is only guaranteed for values high enough of both $L=\sqrt{N}$ and $m$.

\subsection{Evaluation of Density Increasing with the New Set of Solutions}

The density of solutions within the grid affects the final success probability achieved by the LQW algorithm. We already analyzed decreases in the solution density by adding unmarked vertices using both the $M_m$ and $P_{L, m}$ sets. Regarding increases in the density of solutions, we complement this analysis here using the $P_{L, m}$ set since results using the $M_m$ set are found in~\cite{nahimovs_lackadaisical}. Increases in the density of solutions occur by having more marked vertices in a fixed-size grid of $L$ x $L$.

\begin{figure}
    \centering
    \includegraphics[width=0.6\textwidth]{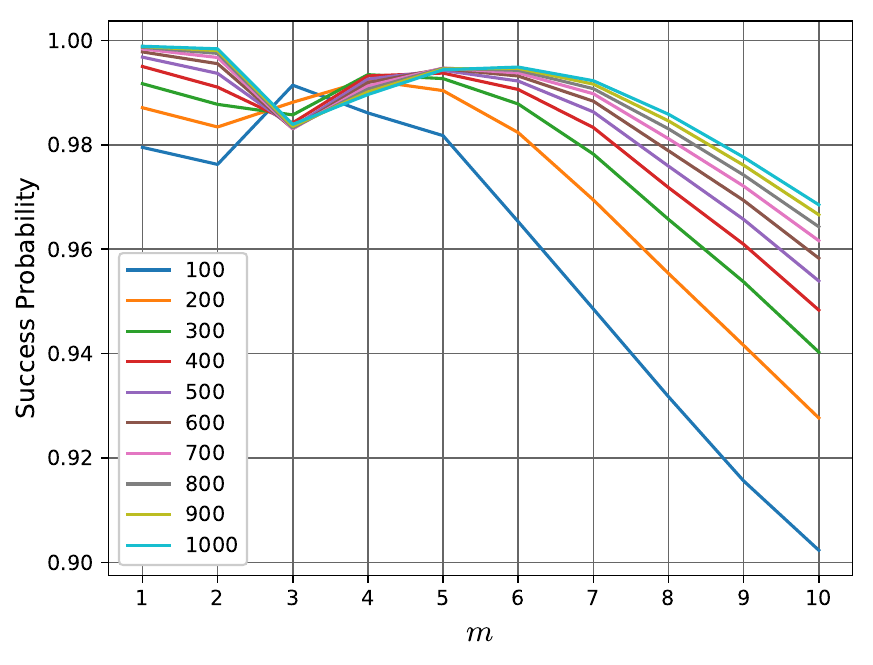}
    \caption{Final success probability as the density of solutions increases with the solutions following the $P_{L, m}$ set. The colored lines represent grids with different numbers $L$ of vertices per dimension.}
    \label{fig:add_m_multiple_L}
\end{figure}

Figure~\ref{fig:add_m_multiple_L} shows the final success probability as a function of the number of solutions $m$ for grids with different numbers $L$ of vertices per dimension. The colored lines represent the results for grids with $L$ varying from $100$ to $1000$ and with the number of solutions $m= \{1,2,\ldots,10\}$ following the $P_{L, m}$ set.

As expected, because of the inversely proportional relation, the success probability decreases as the density of solutions increases by having more solutions in the grid. However, these results with the $P_{L, m}$ set also had that transitory phenomenon. There are intervals where a disturbed behavior exists for all cases, and then the success probability tends to decrease continuously.

This result constitutes one more perspective that shows some uncertainty about the behavior of the LQW algorithm with small values of some input parameters. Thus, it is more confident to apply the LQW search algorithm in real scenarios where the input parameters are higher to avoid all those disturbed behaviors.

%% file: sections/05-lqw_d-dim.tex
\section{Lackadaisical Quantum Walk on d-dimensional Grids}
\label{section:lqw_d-dim_grids}

Marking a vertex as a solution in the two-dimensional case means that its coordinates $x$ and $y$ satisfy the search problem when combined. Searching and retrieving only two values can be a limitation because more than two quantities may be required to solve some applications. The idea here is to expand the technique to search for solutions with an arbitrary number of dimensions, which implies walking through grids with higher than two dimensions. The new capacity is supposed to expand the spectrum of applications of the LQW search algorithm. In the following, the mathematical formalism is redefined, the self-loop weight $l$ is adjusted, and the stopping condition is revisited, all of this considering practical issues that arise in scenarios with arbitrary dimensions.

\subsection{Generalization to $d$-dimensional Grids}

Here, the mathematical formalism is rewritten to the search with the LQW algorithm on grids of higher dimensions. Fortunately, the generalization is straightforward from the two-dimensional formulation, demonstrated in the following algebra. Besides that, we discuss some aspects involving classical simulations of the generalized technique in grids of arbitrary dimensions.

Considering a grid with $d$ dimensions, each vertex has $2d + 1$ possible directions of movement because each dimension has a positive and a negative direction, and there is a self-loop attached to the vertex. In a generalized way, the coin space $\mathcal{H}_C$ is spanned by $\{ \ket{\Uparrow_1}, \ket{\Downarrow_1}, \ldots, \ket{\Uparrow_d}, \ket{\Downarrow_d}, \ket{\circlearrowleft}\}$, where $\Uparrow_i$ and $\Downarrow_i$ represent the movements on the $i$-th dimension. The computational basis for the space of positions $\mathcal{H}_P$ is $\{\ket{x_1,\ldots,x_d} : x_i \in [0,\ldots,\sqrt[d]{N}-1]\}$. Thus, the quantum walk evolves on the space $\mathcal{H} = \mathcal{H}_{C} \otimes \mathcal{H}_{P}$ with these generalized reformulations.

The action of applying $C_1 = -C_0$ to the solutions and $C_0$ to the other vertices can be replaced by an oracle that is used first, followed by $C_0$ acting on all vertices indistinguishably. Since the oracle flips the signs of all $m$ marked vertices, the overall effect is to apply $-C_0$ to the marked vertices and $C_0$ to the others. Using an oracle, $C$ does not need to be broken down into two different coin operators, so $C=C_0$. In a generalized form, the coin operator is now defined as $C = 2 \ket{s_c}\bra{s_c} - I_{2d+1}$, where $\ket{s_{c}}$ is the generalized distribution presented in Equation~\ref{eq:generalized_coin_state}.

\begin{equation}
\label{eq:generalized_coin_state}
\ket{s_c} = \frac{1}{\sqrt{2d+l}} (\ket{\Uparrow_1} + \ket{\Downarrow_1} + \ldots \\
+ \ket{\Uparrow_d} + \ket{\Downarrow_d} + \sqrt{l}\ket{\circlearrowleft})
\end{equation}

For the purpose of applying that coin operator $C$, the outer product $\ket{s_c}\bra{s_c}$ can be written in the matrix form as follows:

\begin{align*}
\ket{s_c}\bra{s_c} &= \frac{1}{\sqrt{2d+l}}
\begin{pmatrix}
1 \\ \vdots \\ 1 \\ \sqrt{l}
\end{pmatrix} \cdot \frac{1}{\sqrt{2d+l}}
\begin{pmatrix}
1 & \ldots & 1 & \sqrt{l}
\end{pmatrix}
\\
&= \frac{1}{2d+l}
\begin{pmatrix}
1 & \ldots & 1 & \sqrt{l} \\
\vdots & \vdots & \vdots & \vdots \\
1 & \ldots & 1 & \sqrt{l} \\
\sqrt{l} & \ldots & \sqrt{l} & l
\end{pmatrix}.
\end{align*}

Let $\ket{\phi}$ be a quantum state of the generalized coin space $\mathcal{H}_{C}$, i.e., $\ket{\phi} = (\alpha_1, \ldots, \alpha_{2d}, \alpha_{2d+1})^T$. Thus, the application of the coin operator $C$ in that generic quantum state is as follows:

\begin{gather*}
(2\ket{s_c}\bra{s_c} - I_{2d+1}) \ket{\phi} =  2\ket{s_c}\bra{s_c}\ket{\phi} - I_{2d+1}\ket{\phi}
\\
= 2 \cdot \frac{1}{2d+l} 
\begin{pmatrix}
1 & \ldots & 1 & \sqrt{l} \\
\vdots & \vdots & \vdots & \vdots \\
1 & \ldots & 1 & \sqrt{l} \\
\sqrt{l} & \ldots & \sqrt{l} & l
\end{pmatrix}
\begin{pmatrix}
\alpha_1 \\ \vdots \\ \alpha_{2d} \\ \alpha_{2d+1}
\end{pmatrix} -
\begin{pmatrix}
1 & \ldots & 0 & 0 \\
\vdots & \vdots & \vdots & \vdots \\
0 & \ldots & 1 & 0 \\
0 & \ldots & 0 & 1
\end{pmatrix}
\begin{pmatrix}
\alpha_1 \\ \vdots \\ \alpha_{2d} \\ \alpha_{2d+1}
\end{pmatrix}
\\
= 2 \cdot \frac{1}{2d+l}
\begin{pmatrix}
\alpha_1 + \ldots + \alpha_{2d} + \sqrt{l} \cdot \alpha_{2d+1} \\
\vdots \\
\alpha_1 + \ldots + \alpha_{2d} + \sqrt{l} \cdot \alpha_{2d+1} \\
\sqrt{l} \cdot (\alpha_1 + \ldots + \alpha_{2d} + \sqrt{l} \cdot \alpha_{2d+1}) \\
\end{pmatrix} - 
\begin{pmatrix}
\alpha_1 \\ \vdots \\ \alpha_{2d} \\ \alpha_{2d+1}
\end{pmatrix}.
\end{gather*}

Defining $\lambda$ as:

\begin{equation*}
\lambda = \frac{1}{2d+l} (\alpha_1 + \ldots + \alpha_{2d} + \sqrt{l} \cdot \alpha_{2d+1}),
\end{equation*}

the application of the coin operator $C$ comes down to:

\begin{equation*}
(2\ket{s_c}\bra{s_c} - I_{2d+1}) \ket{\phi} = 
\begin{pmatrix}
2\lambda - \alpha_1 \\
\vdots \\
2\lambda - \alpha_{2d} \\
2\lambda\sqrt{l} - \alpha_{2d+1}
\end{pmatrix}.
\end{equation*}

It is precisely the result expected when the weighted Grover diffusion coin is applied. Since $\lambda$ is not the mean of the coin amplitudes, the inversion is about a mean that takes into account the weight of the graph edges rather than the simple arithmetic mean~\cite{wong_coin_weighted_graphs}. All this worked out also in the generalized form.

Regarding classical simulations, the coefficients $\alpha_1, \cdots, \alpha_{2d}$ must be handled in a flexible and generalized way according to the total number of dimensions, besides considering the loop weight appropriately to normalize the quantum state. It turns out that the simulation is similar to the one developed by Wong~\cite{wong_lackadaisical_2d}, but considering the generalized relations stated here when it comes to the coin operator.

About the flip-flop shift operator $S_{ff}$, there is also no restriction to generalize it. The intuition behind this operator is to transfer energy between vertices in a dimension-per-dimension way. At a step, what is happening in one dimension for a vertex does not affect what is happening in another dimension for the same vertex. For example, in each step, a vertex $\ket{v}$ stores on its state $\ket{\Uparrow_i}$ the energy coming from the state $\ket{\Downarrow_i}$ of the vertex immediately following on the direction $\Uparrow_i$, and vice versa. Note that, for that dimension, the others do not cause interference.

Therefore, the operator $S_{ff}$ can be generalized by acting on each dimension separately, as presented in Equation~\ref{eq:flip_flop_pattern}. That pattern of dimension-wise energy transfer simplifies an abstraction for a classical simulation. In each step, through a double for-loop, the implementation can replicate the operation in each dimension one by one for each vertex.

\begin{equation}
\label{eq:flip_flop_pattern}
\begin{aligned}
S_{ff} \ket{\Uparrow_i} \ket{x_1,\ldots,x_i,\ldots,x_d} = 
\ket{\Downarrow_i} \ket{x_1,\ldots,x_i+1,\ldots,x_d}
\\
S_{ff} \ket{\Downarrow_i} \ket{x_1,\ldots,x_i,\ldots,x_d} = 
\ket{\Uparrow_i} \ket{x_1,\ldots,x_i-1,\ldots,x_d}
\end{aligned}
\end{equation}

As before, the energy stored in the self-loop remains unchanged after an application of the flip-flop shift operator, i.e., $S_{ff} \ket{\circlearrowleft} \ket{x_1,\ldots,x_i,\ldots,x_d} = \ket{\circlearrowleft} \ket{x_1,\ldots,x_i,\ldots,x_d}$. Thus, no considerations are needed for generalization. As the space topology is torus-like because of the periodic boundary conditions, the shift operates mod $\sqrt[d]{N}$.

Finally, the system begins in the uniform distribution between each of the $N$ vertices of the $d$-dimensional grid with the weighted superposition of coin states generalized in Equation~\ref{eq:generalized_coin_state}. The step where the simulation stops depends on a stopping condition. As we already demonstrated in Section~\ref{section:comparison_stop_conditions}, the more appropriate stopping condition is to monitor the success probability until achieving its maximum. Now, the LQW algorithm can be simulated on higher-than-two-dimensional grids with multiple solutions.

\subsection{Application on $d$-dimensional Grids}

To conduct our experiments on $d$-dimensional grids, we locate the $m$ solutions according to the $P_{d, L, m}$ set presented in Equation~\ref{eq:new_set_solutions_d-dim}, which is a straightforward generalization of the $P_{L, m}$ set already introduced in Equation~\ref{eq:new_set_solutions}. Therefore, each solution is a $d$-tuple, and the solutions are equidistant on the grid's main diagonal.

\begin{equation}
\label{eq:new_set_solutions_d-dim}
    P_{d,L,m} = \Bigg\{ \bigg( \Big\lfloor \frac{L}{m} \Big\rfloor i, \cdots, \Big\lfloor \frac{L}{m} \Big\rfloor i \bigg)_d \ \ \Big| \ \ i \in [0, m-1] \Bigg\}
\end{equation}

Experiments can be performed on grids of higher dimensions since the technique, the stopping condition, and the solution setup are described. The number of steps $T$ and the final success probability $Pr$ for some $d$-dimensional cases are presented in Table~\ref{table:loop_nahimovs_d-dim}. Note that the success probability decreases as the number of dimensions increases, suggesting that the LQW algorithm is ineffective in higher-than-two-dimensional scenarios.

\begin{table}
    \centering
    \caption{Number of steps and final success probability for some cases in grids with higher than two dimensions using $l=\frac{4m}{N}$ and the $P_{d,L,m}$ set.}
    \label{table:loop_nahimovs_d-dim}
    \begin{tabular}{ccccc}
        \hline
        $d$ & $L$ & $m$ & $T$ & $Pr$ \\ \hline
        3 & 32 & 8 & 134 & 0.958805 \\
        4 & 16 & 4 & 257 & 0.888795 \\
        5 & 10 & 5 & 285 & 0.816259 \\
        6 & 8 & 4 & 441 & 0.739591 \\
        \hline
    \end{tabular}
\end{table}

Although $L$ and $m$ varied in the cases presented in Table~\ref{table:loop_nahimovs_d-dim}, the density of solutions was small enough and the relative distance between solutions was high enough to not affect substantially at all. Thus, the final success probability deteriorated strictly due to increases in $d$. Those deteriorated results were found setting $l=\frac{4m}{N}$ for being the optimal $l$ on 2D grids, at least for the best cases of solution densities. It makes room to search for even better adjustments of the self-loop weight since research efforts have already demonstrated how critical adjusting $l$ is~\cite{wong_lackadaisical_2d, nahimovs_lackadaisical}. The following experiment aims to verify whether another optimal value of $l$ for $d$ higher than two exists.

As demonstrated in~\cite{wong_lackadaisical_2d}, $l$ is inversely proportional to the number of vertices $N$. At the same time, $l$ is directly proportional to the number of solutions $m$~\cite{nahimovs_lackadaisical}. Thus, the value of $l$ depends on the density of solutions $\rho$, where $\rho = \frac{m}{N}$. Preserving the relations found by those works, we search for new fits of $l$ in the form $l = \rho \cdot a$, where $a$ is a multiplicative factor. From those previous works, the value of $a$ would be $4$, but we already showed in Table~\ref{table:loop_nahimovs_d-dim} that the LQW deteriorates as $d$ increases with such a value of $a$.

\begin{figure}
    \centering
    \begin{subfigure}{0.49\textwidth}
        \centering
        \includegraphics[width=\textwidth]{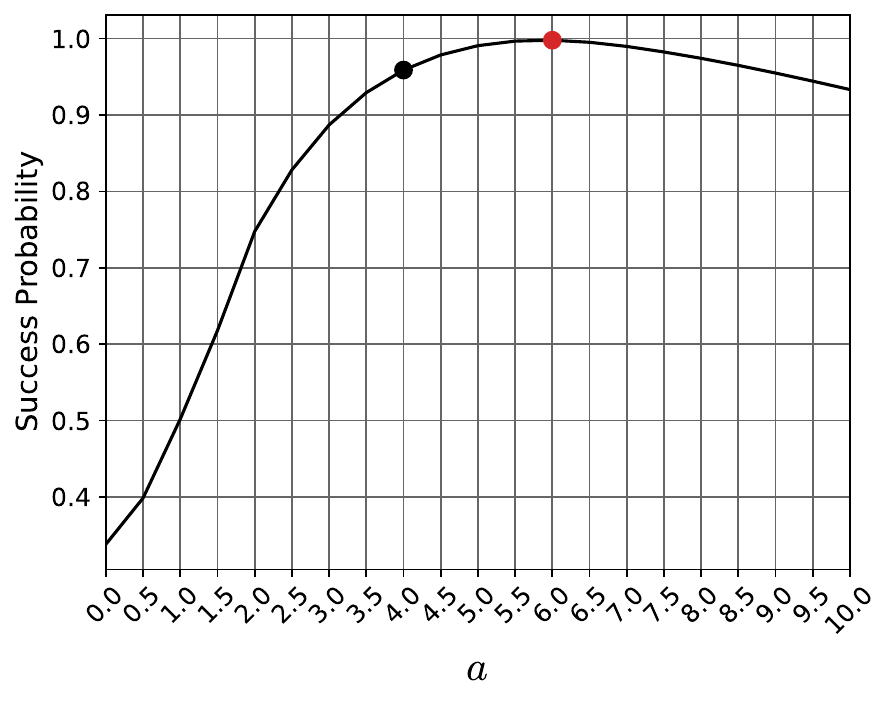}
        \caption{}
        \label{fig:search_loop_3d}
    \end{subfigure}
    \hfill
    \begin{subfigure}{0.49\textwidth}
        \centering
        \includegraphics[width=\textwidth]{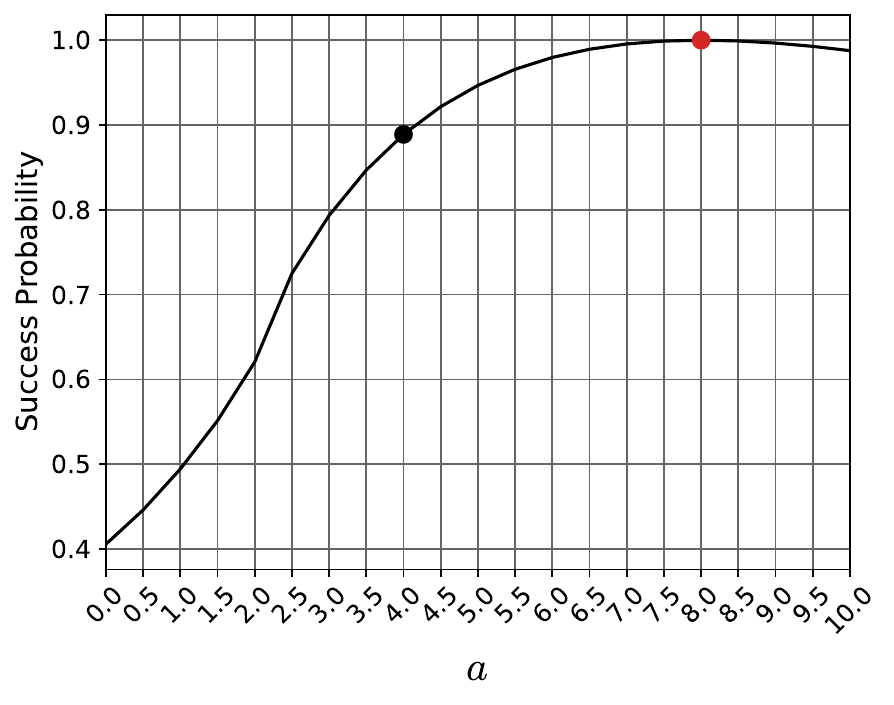}
        \caption{}
        \label{fig:search_loop_4d}
    \end{subfigure}
    \caption{Final success probability as a function of $a$ for the self-loop weight in the form $l=\rho \cdot a$. To the left, results for the 3D case with $L=32$ and $m=8$. To the right, results for the 4D case with $L=16$ and $m=4$. The black dot marks the results for $a$ equal to $4$, while the red dot marks the best overall result found.}
    \label{fig:search_loop_d-dim}
\end{figure}

Figure~\ref{fig:search_loop_3d} and Figure~\ref{fig:search_loop_4d} show the final success probability as a function of that multiplicative factor $a$ for the 3D and 4D cases presented in Table~\ref{table:loop_nahimovs_d-dim}, respectively. The result for $a = 4$ is marked by the black dot, while the red dot marks the best overall result in the search. In both cases, $a = 4$ is not optimal because other values generated better success probabilities. The best overall result of each case could generate success probabilities near $1$, while its distance to the $a = 4$ gets larger when $d$ increases.

Regarding the other cases presented in Table~\ref{table:loop_nahimovs_d-dim}, the best overall results in the search generated success probabilities of $0.999933$ for the 5D case and $0.999986$ for the 6D case. In this way, it is possible to conclude that the LQW algorithm can generate satisfactory results when applied in grids with higher than two dimensions with multiple solutions. The results reported in Table~\ref{table:loop_nahimovs_d-dim} deteriorated because $a$ equal to $4$, proposed by previous works, is optimal only in the restricted 2D case.

Moreover, our results revealed a pattern. The best values of $a$ found in the search were $6$ for the 3D case, $8$ for the 4D case, $10$ for the 5D case, and $12$ for the 6D case. Thus, the experimental results suggest that the optimal value of $a$ is $2 \cdot d$, so the optimal value of $l$ for $d$-dimensional grids is $l = \rho \cdot 2d = \frac{2dm}{N}$. In the 2D case, $l = \rho \cdot 2 \cdot 2 = \frac{4m}{N}$, as proposed in previous works. More experimental evidence is presented in Table~\ref{table:comparison_loops_d-dim}, which compares results obtained using the $l$ proposed in previous works ($l = \frac{4m}{N}$) with the ones obtained using the $l$ proposed in this work ($l = \frac{2dm}{N}$), for a variety of cases in higher than two dimensions.

\begin{table}
    \centering
    \caption{Number of steps and final success probability for some $d$-dimensional grids using the value of $l$ proposed in previous works, $l=\frac{4m}{N}$, and the value proposed in this work, $l=\frac{2dm}{N}$.}
    \label{table:comparison_loops_d-dim}
    \begin{tabular}{ccc|cc|cc}
        \hline
        \multirow{2}{*}{$d$} & \multirow{2}{*}{$L$} & \multirow{2}{*}{$m$} & \multicolumn{2}{c|}{$l=\frac{4m}{N}$} & \multicolumn{2}{c}{$l=\frac{2dm}{N}$} \\ \cline{4-7}
        & & & $T$ & $Pr$ & $T$ & $Pr$ \\ \hline
        3 & 32 & 4 & 187 & 0.959003 & 171 & 0.999531 \\
        3 & 64 & 8 & 381 & 0.959096 & 348 & 0.999736 \\
        4 & 16 & 2 & 364 & 0.888818 & 315 & 0.999912 \\
        4 & 30 & 3 & 1048 & 0.88888 & 907 & 0.99999 \\
        5 & 10 & 2 & 453 & 0.816318 & 377 & 0.999982 \\
        5 & 15 & 5 & 784 & 0.816322 & 658 & 0.999991 \\
        6 & 8 & 2 & 593 & 0.731387 & 600 & 0.999994 \\
        6 & 10 & 10 & 541 & 0.73811 & 525 & 0.999986 \\
        7 & 6 & 6 & 388 & 0.692785 & 354 & 0.99999 \\
        8 & 4 & 2 & 247 & 0.637346 & 295 & 0.999979 \\ \hline
    \end{tabular}
\end{table}

For all cases, $l = \frac{2dm}{N}$ generated success probabilities near to $1$, while $l = \frac{4m}{N}$ generated inferior results that deteriorated further as $d$ increased. For the cases presented in Table~\ref{table:comparison_loops_d-dim}, not only experiments with those two fits of $l$ were made, but no other adjustment surpassed the result of using $l = \frac{2dm}{N}$, indicating that it is the optimal value for $d$-dimensional grids with multiple solutions.

Therefore, this work becomes part of the community efforts that developed the LQW algorithm by adjusting the self-loop weight optimally for different scenarios, as summarized in Table~\ref{table:literature_self-loop_weights}. When there is a single solution in the search space, the optimal self-loop weight is $l=\frac{2}{N}$ for 1D grids~\cite{giri_LQW_runtime_m-solutions} and $l=\frac{4}{N}$ for 2D grids~\cite{wong_lackadaisical_2d}. For vertex-transitive graphs in general, $l=\frac{V}{N}$ is the optimal value, where $V$ is the valency of the graph~\cite{rhodes_LQW_vertex-transitive_graphs}. In contrast, the optimal value is $l=\frac{dm}{N}$ when searching for multiple solutions on $d$-dimensional hypercubes~\cite{luciano_LQW_hypercube_multiple-solutions}.

\begin{table}
    \centering
    \caption{Optimal self-loop weight proposed in different works that developed the LQW algorithm to search distinct scenarios. Basically, the graph structure defines each scenario. The scenarios are also characterized depending on the existence of a single solution or multiple solutions.}
    \label{table:literature_self-loop_weights}
    \begin{tabular}{c|c|c|c}
        \hline
        Work & Graph Structure & Solution & Self-Loop Weight \\ \hline
        Giri and Korepin~\cite{giri_LQW_runtime_m-solutions} & 1D grid & Single & $l=\frac{2}{N}$ \\
        Wong~\cite{wong_lackadaisical_2d} & 2D grid & Single & $l=\frac{4}{N}$ \\
        Rhodes and Wong~\cite{rhodes_LQW_vertex-transitive_graphs} & Vertex-transitive & Single & $l=\frac{V}{N}$ \\
        Souza et al.~\cite{luciano_LQW_hypercube_multiple-solutions} & Hypercube & Multiple & $l=\frac{dm}{N}$ \\
        Saha et al.~\cite{saha_lackadaisical_block_solutions} & 2D grid & Multiple & $l \approx \frac{4}{N(m+1)}$ \\
        Nahimovs~\cite{nahimovs_lackadaisical} & 2D grid & Multiple & $l=\frac{4(m-O(m))}{N}$ \\
        Nahimovs and Santos~\cite{nahimovs_LQW_2D-types} & 2D grid & Multiple & $l=\frac{Vm}{N}$ \\
        This work & $d$D grid & Multiple & $l=\frac{2dm}{N}$ \\
        \hline
    \end{tabular}
\end{table}

Regarding 2D grids again, the adjustment $l \approx \frac{4}{N(m+1)}$ is required if the multiple solutions are arranged as a block~\cite{saha_lackadaisical_block_solutions}. In contrast, $l=\frac{4(m-O(m))}{N}$ is the optimal value for arbitrary placements of the solutions~\cite{nahimovs_lackadaisical}. Searching for multiple solutions on 2D grids with different valencies $V$ requires the value $l=\frac{Vm}{N}$ to achieve optimal results~\cite{nahimovs_LQW_2D-types}. This work, in turn, developed the LQW algorithm to a scenario that was not covered in previous works. The optimal fit of $l$ proposed here, $l=\frac{2dm}{N}$, successfully enables the LQW algorithm to search for multiple solutions on $d$-dimensional grids.

\subsection{Stopping Condition Revisited}

As shown in Section~\ref{section:comparison_stop_conditions}, the more appropriate and natural choice of stopping condition is to monitor the probability evolution about the $m$ marked vertices until this quantity achieves its maximum. The maximum is determined by finding a step whose success probability is smaller than the immediately previous one. That approach assumes a function that increases monotonically, achieves its maximum, and decreases monotonically after that maximum. However, it is not the case in some examples on grids with higher than two dimensions.

Table~\ref{table:comparison_loops_special_cases_d-dim} shows four exceptional cases that stopped at a considerably premature step by using as criterion finding a step whose success probability is smaller than the immediately previous one. We still considered the value of $l$ proposed in previous works ($l=\frac{4m}{N}$), although it is not optimal for $d$-dimensional grids, as well as the optimal value of $l$ found in this work ($l=\frac{2dm}{N}$) to assess that those premature stops are not related with the choice of the self-loop weight. As can be seen, regardless of the value of $l$, each case stopped at the same premature step and, consequently, with a highly unsatisfactory success probability.

\begin{table}
    \centering
    \caption{Number of steps and final success probability for $d$-dimensional cases that prematurely stopped using both the value of $l$ proposed in previous works, $l=\frac{4m}{N}$, and the value proposed in this work, $l=\frac{2dm}{N}$.}
    \label{table:comparison_loops_special_cases_d-dim}
    \begin{tabular}{ccc|cc|cc}
        \hline
        \multirow{2}{*}{$d$} & \multirow{2}{*}{$L$} & \multirow{2}{*}{$m$} & \multicolumn{2}{c|}{$l=\frac{4m}{N}$} & \multicolumn{2}{c}{$l=\frac{2dm}{N}$} \\ \cline{4-7}
        & & & $T$ & $Pr$ & $T$ & $Pr$ \\ \hline
        5 & 10 & 2 & 24 & 0.009348 & 24 & 0.009374 \\
        5 & 15 & 3 & 24 & 0.001847 & 24 & 0.001848 \\
        5 & 15 & 5 & 14 & 0.001108 & 14 & 0.001108 \\
        7 & 6 & 6 & 6 & 0.000878 & 6 & 0.000878 \\ \hline
    \end{tabular}
\end{table}

\begin{figure}
    \centering
    \includegraphics[width=0.6\textwidth]{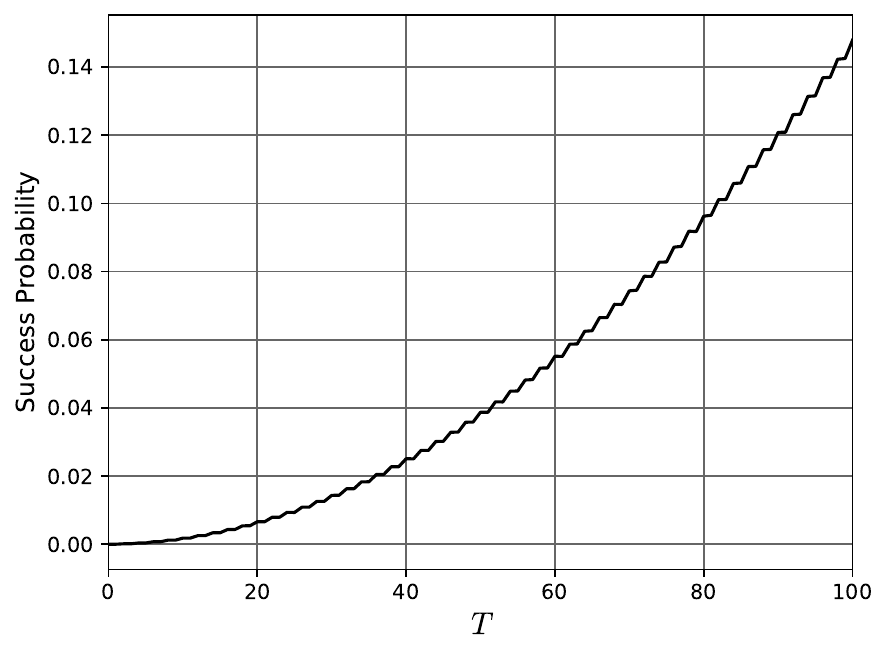}
    \caption{Success probability during the first $100$ steps on the 5D grid with $L=10$, $m=2$, and the value of $l$ proposed in this work, which is $l=\frac{2dm}{N}$.}
    \label{fig:special_case_evolution}
\end{figure}

To evaluate whether or not the system can evolve further, even though the stopping condition is satisfied too early, we stored the success probabilities during $100$ steps of the LQW algorithm with $l=\frac{2dm}{N}$ for the first case presented in Table~\ref{table:comparison_loops_special_cases_d-dim}, which is the 5D grid with $L = 10$ and $m=2$. Figure~\ref{fig:special_case_evolution} shows such a system evolution. Qualitatively, the success probability improves continuously as more steps are performed, but there is a kind of fluctuation in the process. However, this fluctuation has a meaning.

Ambainis et al.~\cite{ambainis_search_walk} mathematically proved that two steps of a particular quantum walk search algorithm give precisely one step of Grover's algorithm. It turns out, in every two steps, the first is an intermediary step to the actual amplitude amplification generated by the second step of the quantum walk. That quantum walk occurred on a complete graph with a non-weighted self-loop for each vertex.

The result we showed in Figure~\ref{fig:special_case_evolution} is supposed to be an experimental demonstration of that two-to-one relation between quantum walks and Grover's algorithm steps. Interestingly, we used a quantum walk with weighted self-loops on grids with higher than two dimensions and not a quantum walk with non-weighted self-loops on a complete graph as used in~\cite{ambainis_search_walk}. Nevertheless, there are similarities because both quantum walk approaches apply a Grover diffusion coin and the flip-flop shift operator $S_{ff}$ subsequently.

The practical implication in terms of the stopping condition is that the success probabilities of adjacent steps must not be compared anymore. As shown explicitly in Figure~\ref{fig:special_case_evolution}, considering two adjacent steps, one of them is an intermediary step subject to fluctuations. Instead of comparing with the immediately previous step, the solution is to compare with the penultimate step.

In this way, the simulation stops in the step whose success probability is smaller than the one of the penultimate step, and the success probability of that penultimate step is reported as the maximum found. This is enough to conceive a more robust stopping condition capable of escaping the premature stops reported in Table~\ref{table:comparison_loops_special_cases_d-dim}. To obtain those results reported in Table~\ref{table:comparison_loops_d-dim}, the stopping condition needs this slight modification, especially for these exceptional cases.

%% file: sections/06-final_remarks.tex
\section{Final Remarks}
\label{section:final_remarks}

This research addressed the LQW search algorithm and its capabilities from an experimental point of view. We aimed to understand properties and existing limitations more clearly, in addition to contributing to a better quantum-walk-based solver of search problems.

In this way, first, we demonstrated that different stopping conditions used in previous works are not interchangeable. Calculating the absolute value of the inner product $\langle \psi (t)|\psi(0) \rangle $ implies prematurely stops. Instead, the real value must be used. After choosing the stopping condition correctly, we demonstrated that the final success probability is inversely proportional to the density of solutions and directly proportional to the relative distance between solutions. However, those relations are guaranteed only for high values of the input parameters. We showed disturbed behaviors in a transition between small to high values of the input parameters from different perspectives.

Consolidating the work, we generalized the LQW algorithm to search for multiple solutions on grids of arbitrary dimensions, not only on the restricted 2D case. However, a new adjustment for the self-loop weight is necessary to obtain successful searches. The experiments we made allow concluding that $l=\frac{2dm}{N}$ is the generalized and optimal value of $l$ for $d$-dimensional grids with multiple solutions. The fits proposed in previous works are only a specific case where $d$ equals $2$. The investigations on $d$-dimensional grids also clarified a two-to-one relation between the steps of the LQW and the ones of Grover's algorithm. An actual amplitude amplification occurs at every two steps, where the first is an intermediary step subject to numerical fluctuations. A fluctuation-tolerant stopping condition is obtained by comparing the success probabilities of the current step and the penultimate step, not between subsequent steps.

Future works should mathematically define upper and lower bounds considering the impacts of multiple solutions stated here. Those impacts of solution densities and relative distances should be studied for solutions randomly sampled from some probability distributions. Another possible direction is to investigate the symmetry breaking~\cite{rapoza_LQW_symmetry-breaking} that nonhomogeneous self-loop weights can cause on grids of arbitrary dimensions with multiple solutions. Inspired by the use of multiple quantum search agents to find optimal solutions for multiobjective optimization problems~\cite{singh_fuzzy-quantum_TS-model}, one more future direction could be to combine the evolution of multiple lackadaisical quantum walkers in the grid. Mathematically or numerically estimated, the number of steps $T$ to the maximum amplitude amplification should be defined \textit{a priori} since it establishes the step where the measurement should occur when executed in quantum devices.

Then, theoretically, the LQW algorithm will be available to execute in quantum devices, ensuring high success probabilities on $d$-dimensional grids with multiple solutions. In practice, the LQW implementation will still need to deal with limitations in the existing quantum hardware. Inspired in~\cite{acasiete_QW_IBM-qdevs}, future works should implement the LQW algorithm on the available quantum computers. Finally, the LQW algorithm should be applied to solve search problems, like the optimization of artificial neural networks, where quantum meta-heuristics of search can be used to tune learning rates~\cite{liu_QGA_for_QNN-tuning}. Moreover, the successful application of the LQW algorithm to transfer quantum states on complete bipartite graphs~\cite{santos_LQW_qstate-transfer_complete-2part-graph} encourages its application for quantum communication on grids.

%% file: sections/07-acknowledgments.tex
\section*{Acknowledgments}

This work was financially supported by the Fundação de Amparo à Ciência e Tecnologia do Estado de Pernambuco (FACEPE), the Conselho Nacional de Desenvolvimento Científico e Tecnológico (CNPq), and the Coordenação de Aperfeiçoamento de Pessoal de Nível Superior - Brasil (CAPES) - Finance Code 001.